  \documentstyle[prb,aps,multicol,epsfig]{revtex}
\begin{document}
% \input epsf.tex
% \tighten
\draft \date{\today} \title
   {Analytic solution for the critical state in
    superconducting elliptic films}
\author{Grigorii P.~Mikitik}
  \address{B.~Verkin Institute for Low Temperature Physics \&
   Engineering, National Ukrainian Academy of Sciences,
   Kharkov 310164, Ukraine}
\author{Ernst Helmut Brandt}
  \address{Max-Planck-Institut f\"ur Metallforschung,
  D-70506 Stuttgart, Germany}
\maketitle

\begin{abstract}
 A thin superconductor platelet with elliptic shape in a
 perpendicular magnetic field is considered. Using a method
 originally applied to circular disks, we obtain an approximate
 analytic solution for the two-dimensional critical state of this
 ellipse. In the limits of the circular disk and the long strip
 this solution is exact, i.e.\ the current density is constant in
 the region penetrated by flux. For ellipses with arbitrary
 axis ratio the obtained current density is constant to typically
 $10^{-3}$, and the magnetic moment deviates by less than $10^{-3}$
 from the exact value. This analytic solution is thus very accurate.
 In increasing applied magnetic field, the penetrating flux fronts
 are approximately concentric ellipses whose
 axis ratio $b/a \le 1$ decreases and shrinks to zero when
 the flux front reaches the center, the long axis staying finite
 in the fully penetrated state. Analytic expressions for these
 axes, the sheet current, the magnetic moment, and the perpendicular
 magnetic field are presented and discussed.
 This solution applies also to superconductors with anisotropic
 critical current if the anisotropy has a particular, rather
 realistic form.
 \end{abstract}
    \begin{multicols}{2}   % \twocolumn
    \narrowtext

\section{Introduction}  % 1

 Most  experiments with high-$T_c$ superconductors deal with thin flat
 samples in a perpendicular magnetic field $H$, for example
 $c$-axis-oriented monocristalline platelets or films. In this
 connection, in recent years the problem of the critical state of
 such samples has attracted considerable interest, see
 e.g.\ Ref.\ \onlinecite{1} and the references cited therein. Exact
 analytic solutions were obtained only for a circular disk \cite{2}
 and a thin infinitely long strip.\cite{3,4,5} In deriving these
 solutions it was essential that the critical states of the disk and
 the strip have a known symmetry. For the strip the critical state
 distributions of the magnetic field and the current do not depend
 on the longitudinal coordinate while the disk has a rotational axis.
 In both cases the directions of circulating currents are
 fixed and known in advance, and thus the critical state equations are
 one-dimensional (1D). In the present paper we obtain an approximate
 analytic solution of a two-dimensional (2D) problem, namely, a
 thin superconductor platelet of elliptic shape in a perpendicular
 magnetic field. This solution enables one to understand the
 critical states of real superconductors when the above-mentioned
 symmetry is absent. The accuracy of our solution is very high,
 especially for the integral quantities like the magnetization
 of the sample, and it may be used to analyze experimental data.
 Besides this, the analysis of our new solution reveals novel
 qualitative features of the critical state of thin superconductors
 like flux fronts with varying curvature and the possibility of
 rotating flux-line arrangments.

 We derive this solution in Sec.\ II and analyze it in Sec.\ III,
 reproducing the known solutions for disks and strips and discussing
 the interesting new features. In Sec.\ IV we apply our solution to
 anisotropic superconductors,
 and in Sec.\ V we summarize the results.

  \section{Derivation}  % 2

 Let us place the origin of the coordinate system in the center of the
 sample and let its plane coincide with the $xy$ plane. The external
 magnetic field $H$ is directed along the $z$ axis, i.e., along the
 thickness $d$ of the sample. The boundary of the  superconductor in
 the $xy$ plane, $\Gamma$, is described by the ellipse (see Fig.\ 1)
   \begin{eqnarray}
   {x^2 \over a_0^2} + {y^2 \over b_0^2} = 1 \, \nonumber
   \end{eqnarray}
 where $a_0$  and $b_0$  are the semi-axes of the ellipse
 ($a_0 \ge b_0 \gg d$).  The critical current density $j_c$ is assumed
 to be constant. The critical-state equations for the
 thickness-integrated current density
 ${\bf J}(x,y) = \int_{-d/2}^{d/2} {\bf j}(x,y,z) dz$ in the partly
 penetrated  critical state have the following form. In the region
 between $\Gamma$ and the penetrating flux front $\gamma$ one has
  \begin{eqnarray} % 1
   | {\bf J}(x,y) | = j_c d \equiv J_c \,.
   \end{eqnarray}
 Here the common assumption has been made that the direction of the
 current does not depend on the coordinate $z$ across the small
 thickness of the sample. In the flux-free region,
 i.e.\ inside $\gamma$, the
 perpendicular induction should vanish,
   \begin{eqnarray}  % 2
    B_z(x,y) = 0  \,.
   \end{eqnarray}
 In addition, the equation
   \begin{eqnarray}  % 3
   \nabla \cdot {\bf J}(x,y) = 0
   \end{eqnarray}
 must hold for any $x$ and $y$ inside the sample.
 In deriving the solution we follow the method of Mikheenko and
 Kuzovlev.\cite{2} When $\gamma$ coincides with $\Gamma$,
 i.e., the flux does not penetrate the superconductor, the solution
 of Eqs.\ (2) and (3) may be found by considering the field on the
 surface of an ideally screening oblate ellipsoid in a
 uniform magnetic field $H \| z$. The field at the surface of the
 ellipsoid is tangential to it and has the form \cite{6}
   \begin{eqnarray}
   {\bf H}_t (x,y,z) = {H \over 1-N_{zz} } \Big({\bf \hat z -
   \hat n( \hat z \hat n)} \Big) \,.  \nonumber
   \end{eqnarray}
 Here  ${\bf\hat z}$  is the unit vector along $z$, ${\bf \hat n}$
 is the unit vector normal to the surface, and $N_{zz}$ is the
 appropriate demagnetizing factor.  For a thin oblate ellipsoid with
 semi-axes $a_0\ge b_0 \gg c_0$  one has \cite{7}
 $1-N_{zz} = (c_0/b_0)E(k)$ where $E(k)$ is the complete elliptic
 integral of the second kind with $k^2 = 1-b_0^2/a_0^2$. For
 example, when $a_0 =b_0$ (disk) one has $E(0)= \pi/2$ and
 $1-N_{zz} = \pi c_0/2a_0$, and when $a_0 \gg b_0$ (strip)
 one has $E(1) = 1$ and $1-N_{zz} = c_0/b_0$.
 Letting  $c_0 \to 0$ and taking into account the relation
 between the surface screening current and the tangential field,
 ${\bf J}_{\rm surf} = {\bf \hat n \times H}_t$, \cite{6} we
 arrive at the sheet current
   \begin{eqnarray}  \nonumber
   {\bf J}(x,y) =  2\, {\bf \hat z \times H}_t   \,,
   \end{eqnarray}
 where ${\bf H}_t$ is the field on the upper side of the platelet.
 This result applies to films of any shape; it means that the
 magnitude of the field change across the film thickness equals $J$.
 In thin ellipses, in the region $f(x/a_0 ,y/b_0 ) <1$ with
 $f(u,v) \equiv u^2 +v^2$, one obtains for  ${\bf J} = (J_x, J_y)$:
   \begin{eqnarray} % 4
   J_x= {2H \over E(k) }~ {y\over b_0} ~   S \equiv -H F_x\left(
      {x\over a_0}, {y \over b_0}, {b_0 \over a_0} \right) \,,
          \nonumber \\
   J_y=-{2H \over E(k) }\, {xb_0\over a_0^2} S \equiv -H F_y\left(
      {x\over a_0}, {y \over b_0}, {b_0 \over a_0} \right) \,,
   \end{eqnarray}
 where $S = 1/\sqrt{1 -x^2/a_0^2 - y^2/b_0^2}$ and
 $k^2  =1 - (b_0 /a_0 )^2$.
 If $f(x/a_0, y/b_0 ) > 1$, we define $F_x  =F_y  =0$. When a
 partially penetrated critical state occurs in the sample, we look
 for the solution of Eqs.\ (1), (2), (3) as a linear
 combination of distributions like that described by formulae (4). In
 other words, we assume that the front of flux penetration has the
 shape of an ellipse with the ratio of the semi-axes depending on the
 depth of the penetration, see Fig.\ 1. This assumption is exact for
 disks and strips and also in the limiting case of a small depth of
 penetration. As will be seen from the results obtained below, the
 deviation from this assumption increases with increasing depth but
 remains very small. The proposed approximation thus
 is sufficient for most practical purposes.

 Let the penetrating flux front be the ellipse
   \begin{eqnarray} % 5
   f \left({x \over a(b)}, {y \over b} \right) = 1
   \end{eqnarray}
 with the long axis $a$ depending on the short axis $b$. Then, according
 to the above assumption, we write for the components $J_x$ and $J_y$
 of the sheet current,
   \begin{eqnarray} % 6
   J_{x,y} = -H \int_b^{b_0} \! W(b')\, F_{x,y} \left(
    {x\over a(b')}, {y\over b'}, {b'\over a(b')} \right) \, db' \,,
   \end{eqnarray}
 where $W(b')$ is a weight function and $F_x$ and $F_y$ are defined
 by Eqs.\ (4). The sheet current (6) with any weight $W(b')$
 satisfies  Eq.\ (3) due to the definition of the functions $F_x$ and $F_y$.
 In the elliptic region
   \begin{eqnarray}  \nonumber
   f \left({x \over a(b)}, {y \over b} \right) < 1
   \end{eqnarray}
 this current creates a constant field equal to
   \begin{eqnarray}  \nonumber
    -H \int_b^{b_0} \! W(b')\, db' \,,
   \end{eqnarray}
 which has to compensate the applied field $H$.
 Thus, condition (2) reduces to the equation
   \begin{eqnarray} % 7
     \int_b^{b_0} \! W(b')\, db' = 1 \,.
   \end{eqnarray}
 The functions $W(b )$ and $a(b)$ must be obtained from the equality
  \begin{eqnarray} % 8
   J_x^2 +J_y^2 = J_c^2
  \end{eqnarray}
 that follows from Eq.\ (1). Let $x=0$. Then Eq.\ (8) takes the form
 which was considered in Ref.\ \onlinecite{2}, and we find
  \begin{eqnarray} % 9
   W(b) = {J_c \over \pi H} {E(k) \over b \sqrt{1 -b^2 /b_0^2} }
  \end{eqnarray}
 with $k^2 =1 - [b /a(b)]^2 $. Now we can obtain the function
 $J_x (x=0,y)$ [while $J_y(x=0,y)=0$] and calculate the total
 current $I$ circulating in the sample,
  \begin{eqnarray} % 10
   I = \int_0^{b_0} \!\! J_x(0,y)\, dy = {2\over \pi} J_c b_0
     \arccos {b \over b_0} \,.
  \end{eqnarray}
 The $H$ dependence of the penetration depth $b(H)$ along the
 $y$ axis is determined by Eq.\ (7). This yields,
   \begin{eqnarray} % 11
    {H \over H_c} = \int_b^{b_0} \! { E(k) \, db' \over
     b' \sqrt{1-b'^2 /b_0^2} } \,.
   \end{eqnarray}
 with $H_c \equiv J_c /\pi \equiv j_c d /\pi$ and
 $k^2 = 1 -[b'/a(b')]^2$, see Fig.\ 2.
 To determine the dependence $a(b)$ let us repeat the above
 reasoning, Eqs.\ (5) to (10), but now we shall consider $b$ as
 a function of $a$ and integrate over $a'$ from $a$ to $a_0$.
 At $y=0$ we can find $W(a')$ and calculate $I$ again, obtaining
  \begin{eqnarray} % 12
   I = -\int_0^{a_0} \!\! J_y(x,0)\, dx = {2\over \pi} J_c a_0
     \arccos {a \over a_0} \,.
  \end{eqnarray}
 Since the total current crossing any radius must be the same, we
 may equate the currents (10) and (12). This yields the dependence of
 $a$ on $b$ (Fig.\ 3),
   \begin{eqnarray} % 13
   a(b) =a_0 \cos \left({b_0\over a_0} \arccos {b\over b_0} \right)\,.
   \end{eqnarray}
 Formulae (4)--(6), (9), (11), and (13) give the solution for the
 critical state in the ellipse. If this solution were exact,
 Eq.\ (8) would hold for any $x$ and $y$ in the region between the
 curves $\Gamma$ and $\gamma$. The numerical analysis shows that, in
 reality, the quantity $\tilde J \equiv ({J_x^2 +J_y^2})^{1/2}/J_c$
 gradually deviates from 1 as the point $(x,y)$ moves away from
 $\Gamma$.  However, this deviation remains small, typically
 $10^{-4} \cdots 10^{-2}$, and is not visible in  Figs.\ 4 - 7.
 In the fully or almost fully  penetrated state near the points
 $x= \pm a(b=0)$, $y=0$, this deviation may reach a few percent.
 Our expressions provide thus a
 very good description of the critical state in elliptic platelets.
 In particular, since this small deviation of $J(x,y)$  from $J_c$
 occurs mainly near the axis $y=0$, its influence on the magnetic
 moment will be negligible, see Sec.\ III.

   Interestingly, an explicit expression can be obtained for the
 $b$-dependence of the magnetic moment $m$ of our elliptic disk
 in the critical state. In the ideal screening state any ellipsoid
 with volume $V$ and semi-axes $a_0, b_0, c_0$ in a field $H \| z$
 has the magnetic moment
  \begin{eqnarray}   \nonumber
    m_{\rm ideal}(a_0, b_0, c_0) = - {VH \over 1 -N_{zz}} =
    - {4\pi \over 3} {a_0 b_0 c_0 \over 1 -N_{zz}} \, H \,.
   \end{eqnarray}
 For thin oblate ellipsoids with $a_0 \ge b_0 \gg c_0$ the
 demagnetizing factor is \cite{7}
 $N_{zz} = 1- (c_0/b_0)E(k)$ with $k^2 = 1-b_0^2/a_0^2$.
 Taking the limit $c_0 \to 0$ one obtains the magnetic moment
 of an ideally screening thin ellipse,
  \begin{eqnarray}   % 14
    m_{\rm ideal}(a_0, b_0) = -{4\pi \over 3}
    {a_0 b_0^2 \over E(k)} \,H  \,.
  \end{eqnarray}
 For the thin ellipse in the critical state we write
    \begin{eqnarray}   % 15
    m = \int_b^{b_0}\!\! W(b') \, m_{\rm ideal}[a(b'),b'] \,db'\,.
  \end{eqnarray}
 With the weight (9) the elliptic integral $E(k)$ drops out and
 this integral yields
   \begin{eqnarray} % 16
   m = -{2 \over 3} J_c a_0 b_0^2 \left[ { \cos(\arcsin \tilde b
   +e\, \arccos \tilde b ) \over 1-e } \right. \nonumber \\ + \left.
   { \cos(\arcsin \tilde b -e\, \arccos \tilde b ) \over 1+e }
           \right] \,,
   \end{eqnarray}
 where $e=b_0 /a_0$ and $\tilde b=b/b_0$. The dependence $m(H)$
 is obtained from Eqs.\ (11) and (16), see Fig.\ 8.

   Next we give the expression for the magnetic field $H_z(x,y)$
 in the penetrated region between $\Gamma$  and $\gamma$.
 Let the point $(x,y)$ be located on the ellipse
   \begin{eqnarray}  % 17
      {x^2\over a^2_{xy}}+ {y^2\over b^2_{xy} } = 1
   \end{eqnarray}
 with semi-axes $a_{xy}$ and $b_{xy}$, where $b \le b_{xy} \le b_0$
 and $a_{xy} = a(b_{xy})$ satisfies Eq.\ (13), i.e., the ellipse is
 a previous flux front, Fig.\ 1. Then we obtain
   \begin{eqnarray} % 18
    H_z(x,y) = H_c \int_b^{b_{xy}}\!\!
    { db' \over \sqrt{1- b'^2 /b_0^2}}
    \nonumber \\ \times \left\{ {b'\over a(b')} \sqrt{\xi +a^2(b')
    \over \xi(\xi +b'^2) } + { E(\varphi,k) \over b'} \right\} \,.
   \end{eqnarray}
 Here $E(\varphi ,k)$ is the elliptic integral of the second kind,
 $\sin\varphi =[\xi /(\xi +b'^2)]^{1/2}$, $k^2 =1-[b'/a(b')]^2$, and
   \begin{eqnarray} \nonumber
   2\xi = x^2\! +y^2\! -a'^2\! -b'^2 \! +        % \\ \nonumber
   \Bigl[ \big( x^2\! +y^2\! -a'^2\! -b'^2 \big)^2   \\ \nonumber
   -\,4 \big( b'^2 a'^2\! - x^2 b'^2\! - y^2 a'^2 \big) \Bigr]^{1/2}
   \end{eqnarray}
 with $a' = a(b')$ from Eq.\ (13). The magnetic field
 outside the superconductor in the $z=0$ plane is given by the same
 expression (18) but with the upper boundary  $b_{xy}$ replaced by
 $b_0$, and inside the flux front $\gamma$ one has $H_z = 0$,
 see Appendix A. Contour lines and profiles of $H_z(x,y)$
 are depicted in Figs.\ 9 - 11.
 At large distances $r=(x^2+y^2)^{1/2} \gg a_0 \ge b_0$ the
 field (18) becomes
 $$ H_z(x,y) \to H - m/(4\pi r^3) \,, $$
    as expected,\cite{6} with magnetic moment $m$ from Eq.\ (16).

    To end this section we give an elegant representation of the
 sheet current ${\bf J}(x,y)$. Substituting Eqs.\ (4) and (9)
 into Eq.\ (6), the expression for  the components $J_x$  and $J_y$
 can be written as follows,
   \begin{eqnarray}  % 19
      J_x =  J_c {\partial g \over \partial y} , ~~
      J_y = -J_c {\partial g \over \partial x} \,, \nonumber \\
%   \end{eqnarray}
% where
%   \begin{eqnarray}
    g(x,y) = - {2\over \pi} \int_{b_{xy}}^{b_0}\!\! { db' \over
    \sqrt{1- b'^2 /b_0^2} } \sqrt{ 1- {x^2 \over a^2(b')} -
     {y^2 \over b'^2 } } \,.
   \end{eqnarray}
 As in Eq.\ (18) the boundary $b_{xy}$ is the semi-axis of a
 previous flux front defined by Eq.\ (17) if the point $(x,y)$ is
 located between the curves  $\Gamma$ and $\gamma$, and $b_{xy} =b$
 when the point is inside $\gamma$ (in the field-free core). For points
 outside the superconductor we put $b_{xy} = b_0$, yielding $g=0$.
   The function $g(x,y)$ is the local magnetization (or density
 of current loops) introduced in Ref.\ \onlinecite{8}; the contour
 lines  $g(x,y)=$ const coincide with the current stream lines;
 one has $g(x,y) = 0$ on the edge $\Gamma$ of the elliptic sample
 (and outside), and the volume under the ``mountain'' $g(x,y)$,
 coincides with the magnetic moment (16).
 We emphasize that the current stream lines in the general ellipse
 are {\it not parallel} to the flux front $\gamma$, see Fig.\ 6,
 in contrast to the situation in thin disks and strips, and
 in longitudinal geometry.

 \section{Analysis}  % 3

    One easily verifies that the obtained expressions for the
 penetration depth, Eq.\ (11), the magnetic field, Eq.\ (18),
 and the currents, Eq.\ (19), go over into the appropriate formulae
 for the disk \cite{2} when $a_0 = b_0$ and for the strip \cite{3,4,5}
 when $a_0 \gg b_0$. For the strip one has $E(k=1)=1$ and formula
 (11) yields the flux-front position
   \begin{eqnarray}  % 20
    {b \over b_0} = {1\over \cosh (H / H_c) } \,,
   \end{eqnarray}
 which coincides with Eq.\ (3.5) of Ref.\ \onlinecite{4} since
 $H_c = J_c /\pi$. For the disk, $E(k=0) = \pi /2$ and we arrive
 at the same expression (20) but with $H_c$ substituted by
 $\pi H_c /2 = J_c /2$, see Eq.\ (19)  of Ref.\ \onlinecite{2}.
 For general ellipses, $b(H)$ falls between the curves for the
 disk and the strip, see Fig.\ 2. It should be noted that the
 penetration along the $x$ axis does not go all to the center
 but $a(b)$ (13) approaches the finite value
 $a(0)= a_0 \cos(\pi e /2)$  when $b\to 0$.

   As for the magnetic moment, when $a_0 = b_0$
 Eq.\ (16) goes over to the appropriate formula for disks,\cite{2}
   \begin{eqnarray} % 21
  m_{\rm disk}  = - {2\over 3} J_c a_0^3  \Big(\! \arccos
  {1\over \cosh h} +{\sinh |h| \over \cosh^2 h}\, \Big)
   \end{eqnarray}
 with $h=2 H /J_c$. In the limit $a_0 \gg b_0$,
 Eq.\ (16) yields the magnetic moment of strips with width $2w$ and
 length $2a_0$,
   \begin{eqnarray} % 22
  m_{\rm strip} = -2 J_c a_0 w^2  \tanh (H/H_c) \,,
   \end{eqnarray}
 with an effective width $2w=(2/3)^{1/2} 2b_0$. This $m_{\rm strip}$
 deviates
 from the result for a strip with width $2b_0$ by a factor $2/3$.
 The reduced effective width is explained as follows. A long
 narrow ellipse with semi-axes $a_0 \gg b_0$ at each position
 $x$ may be approximated by a piece of a long strip with half width
 $y(x)$, having a magnetic moment proportional to
 $y^2(x)/ b_0^2 = 1 - x^2 /a_0^2 $, cf.\ Eq.\ (22). Averaging this
 prefactor over the length $2a_0$ of the ellipse one arrives at the
 reduction factor $2/3$.

    The magnetization curves $m(H)$ for disks (21), strips (22),
 and the general ellipse, Eqs.\ (16) and (11), coincide almost
 exactly when they are normalized such that their initial slope and
 saturation value are unity. The deviation of these reduced curves
 $M_r(H_r)$ from the strip result $\tanh(H_r)$ are always smaller
 than 0.013 as shown in Fig.\ 8. The maximum deviation occurs for
 ellipses with excentricity $e=b_0/a_0 \approx 0.7$. Taking the
 initial slope $m'(0) = \partial m/ \partial H |_{H=0}$ from
 Eq.\ (14),
  \begin{eqnarray}   % 23
    m'(0) = {m_{\rm ideal} \over H} =
   -{4\pi \over 3} {a_0 b_0^2 \over E(k)} \,,
  \end{eqnarray}
 ($k^2 = 1-e^2$, $e=b_0/a_0$), and the saturation value
 $m_{\rm sat} = m(H \to \infty)$ from Eq.\ (16),
   \begin{eqnarray} % 24
   m_{\rm sat} = m(b \to 0) = -{4 \over 3} J_c a_0 b_0^2 \,
   {\cos(e\pi/2) \over 1-e^2} \,,
   \end{eqnarray}
 and approximating the normalized curves by $M_r(H_r) = \tanh(H_r)$
 with $M_r = m /m_{\rm sat}$ and $H_r = H/H_1$,
   \begin{eqnarray} % 25
   H_1 = {m_{\rm sat} \over m'(0)} = J_c {E(k) \over \pi}
   {\cos(e\pi/2) \over 1-e^2} \,,
   \end{eqnarray}
 we get an excellent approximation for the magnetic moment of
 thin ellipses in the Bean model,
   \begin{eqnarray} % 26
   m(H) \approx m_{\rm sat} \tanh(H/H_1) \,,
   \end{eqnarray}
 with $m_{\rm sat}$ and $H_1$ from Eqs.\ (24) and (25). Formula (26)
 is good also for rectangular films, where $m_{\rm sat}$ is well
 known and $m'(0)$ has to be obtained numerically using the method
 of Ref.\ \onlinecite{9}.

  The sheet current ${\bf J}$, Eqs.\ (6) and (19), reduces to
 the known results for circular disks \cite{2} and infinite
 strips. \cite{3,4,5}  In both limits one obtains the same functional
 form for the magnitudes $J(r) = J_c f(r)$ and  $J(y) = J_c f(y)$,
 with the same function:  $f(y) = 1$ for $b \le |y| \le b_0$ and
   \begin{eqnarray} % 27
    f(y) = {2\over \pi}\arctan{cy\over \sqrt{b^2 -y^2}}
   \end{eqnarray}
 for $|y| \le b$, where $c = (1-b^2/b_0^2)^{1/2} = \tanh(h)$ with
 $h=2H/J_c$ for the disk and $h=\pi H/J_c$ for the strip.
 For long ellipses with finite length $2a_0$ in these expressions
 $b_0$ should be replaced by the effective half width $y(x)$ as
 discussed after Eq.\ (22).

   There is another known result with which our solution may be
 compared. Namely, in the flux-penetrated region where $J=J_c$ is
 constant, the current stream lines are equidistant and are thus the
 ``distance function''.\cite{10} This means all points on a given
 stream line have the same distance from the specimen edge. This
 property applies to thin films of {\it any} shape. It applies also
 in the longitudinal geometry, i.e., in both limits of small and
 large thickness. (In the non-penetrated region the current stream
 lines are different in these two limits: in the longitudinal limit
 one has there $j \equiv 0$ and $J\equiv 0$, but in the transverse
 limit $J\ne 0$).
 The current stream lines in the penetrated region are thus the
 envelope lines of circles with constant radius $r$
 ($0 \le r \le b_0-b$) and with center on the specimen edge.
 Parametrizing our elliptic edge by
 $x(\varphi) = a_0 \cos \varphi$, $y(\varphi) = b_0 \sin \varphi$,
 we find for these envelopes
   \begin{eqnarray} % 28
    x(\varphi, r) = (a_0 - r b_0 /W) \cos\varphi \,, \nonumber \\
    y(\varphi, r) = (b_0 - r a_0 /W) \sin\varphi \,,
   \end{eqnarray}
 with $W = (b_0^2 \cos^2\varphi + a_0^2 \sin^2\varphi)^{1/2}$.
 The exact stream lines (28) have a  bend (discontinuity in their
 slope) in the penetrated region on the $x$ axis at
 $a(b) < |x| < a_0 -b_0^2/a_0$,
 but they are smooth at $ a_0 -b_0/a_0^2 \le |x| \le a_0$.
 Our approximate solution
 coincides with these exact stream lines within line thickness.

      From the stream lines in the critical state, Eq.\ (28), one
 obtains the exact magnetic moment of thin ellipses in the
 fully penetrated critical state,
   \begin{eqnarray} % 29
   m_{\rm sat}^{\rm exact} = -{4 \over 3} J_c a_0 b_0^2 \,
   [\, E(k) - {e^2 \over 2} K(k) \,] \,,
   \end{eqnarray}
 where $K(k)$ and $E(k)$ are the elliptic integrals of the first
 and second kind with $k^2 = 1-e^2$, $e=b_0/a_0$, see Appendix A.
 Comparing the exact result (29) with our approximate $m_{\rm sat}$,
  Eq.\ (24), we find that the deviation is extremely small:

 Writing $f_1(e)= \cos(e\pi/2)/(1-e^2)$ and $f_2(e)=E(k) -(e^2/2)K(k)$
 we find $f_1(1) =f_2(1) =\pi/4$ (disk), $f_1(0) = f_2(0) = 1$
 (strip), and $0 \le f_1 - f_2 \le 0.001317$. The maximum deviation
 $f_1-f_2 = 0.001317$ occurs at $e=0.5043$. The deviation is an almost
 symmetric function of the excentricity $e$,
 $f_1(e)-f_2(e) \approx 0.00138 \sin^2(e\pi)$.
 With Eq.\ (29) the field $H_1$ (25) which enters formula (26)
 takes the exact value
   \begin{eqnarray} % 30
   H_1^{\rm exact} = {m_{\rm sat}^{\rm exact} \over m'(0)} =
   H_c E(k) [\, E(k) -  {e^2 \over 2} K(k) \,] \,.
   \end{eqnarray}

    The flux fronts, current stream lines, and contour lines of the
 magnetic field $H_z(x,y)$ in the general elliptic film do not
 coincide, as opposed to the situation in circular disks, long strips,
 and in longitudinal geometry. These features of the critical state
 in non-circular thin  superconductors of finite size are seen even
 more clearly in square and rectangular films,  \cite{9,10,11}  but
 while those geometries were solved numerically, we now have an
 analytic solution for the elliptic film .

    Another novel feature of the critical state
 in thin elliptic films as compared to the
 critical state in circular disks and infinite strips,
 is that the {\it direction} of the sheet current
 ${\bf J}$ at a given point  $(x,y)$ {\it changes} during the
 penetration of flux, see Fig.\ 7. This fact may have an
 important physical consequence for films or platelets with
 thickness  exceeding the London penetration depth $\lambda$.
 As is known,\cite{12,13,14} in flat superconductors with
 $d\gg \lambda$ in moderate fields $H$ there is a flux and
 current-free core described by $ |z| < z_{\rm core}(x,y)$.
 In the cases of strips and disks
 this core was depicted, e.g., in Refs.\ \onlinecite{15,16}.
 Critical currents circulate only outside
 the core, for $ z_{\rm core}(x,y) < |z| < d/2$. In this shell the
 critical state has the usual Bean form,  but the field gradient
 occurs {\it across the thickness} of the sample since this state is
 forced by the screening currents, and the flux lines are almost
 {\it parallel} to the flat surface.
 The outer rim (equator) of the flux-free core coincides with the
 penetrating flux front since in the region inside the flux front
 the perpendicular magnetic field $B_z$ is practically zero.

   Now, since the tangential field at the flat surface is
 perpendicular to the sheet current ${\bf J}$, the changing
 direction of ${\bf J}(x,y; b)$ during flux penetration (i.e.\ with
 increasing $H$ and decreasing $b$) means that the direction of
 the penetrating flux lines in the region below $\gamma $ also
 changes. Thus,  flux lines {\it with gradually rotating orientation}
 enter through the flat surface at each point away from the
 symmetry axes. These U-shaped flux lines move towards the specimen
 center, dragged by the Lorentz force, which is balanced by bulk
 pinning. As a result, in the region inside $\gamma$ the layers
 of the flux-line lattice exhibit torsion relative to each
 other and therefore carry a longitudinal current component,
 i.e., the local current density  is not exactly perpendicular to
 the flux lines.
 In principle, this may lead to instabilities and flux-cutting
 processes. \cite{17,18,19}  That situation may be related to the
 penetration of a rotating field component discussed by
 Bean \cite{20} and by Gilchrist.\cite{21}

   In deriving the solution we have assumed that
 ${\bf B}=\mu_0 {\bf H}$ and the so called geometrical barrier
 \cite{22,23,24,25} is negligible. These assumptions hold if the
 characteristic magnetic field $j_c d$ considerably exceeds the
 lower critical field $H_{c1}$. In the case $j_c d < H_{c1}$ and
 $H < H_{c1}$ the critical state problem becomes more
 complicated. Analytical  \cite{22,23} and numerical \cite{24,25}
 methods to analyze this general problem were recently proposed.
 In particular, if our assumption
 ${\bf B}=\mu_0 {\bf H}$ is replaced by the correct $H(B)$ (obtained,
 e.g., from Ginzburg-Landau theory \cite{26}) then flux does
 not penetrate until the applied field $H$ has reached the field
 of first penetration of flux,
 $H_{en} \approx H_{c1} \tanh\sqrt{c d/2w}$,
 where $c\approx 0.36$ for strips with half width $w$ and
 $c \approx 0.67$ for disks with radius $w$. \cite{25}

\section{Anisotropic Critical Current}  % 4

The obtained solution allows one to analyze the critical state of
flat superconductors when the critical current density  $j_c$
is {\it anisotropic} in the plane of the film. Such anisotropy can
occur, for example, if there are twin boundaries or other extended
or inclined defects, or if the superconductor itself is anisotropic.
A controlled in-plane anisotropy of $j_c$ of more complicated nature
may be induced by applying a magnetic field parallel to the
film.\cite{27}  To explain our analysis, we perform a
transformation of coordinates:
   \begin{eqnarray} % 31
     x = \alpha x', ~~~ y= \beta y' \,,
   \end{eqnarray}
where $\alpha $ and $\beta $ are some constants and the prime
denotes the transformed quantities. Under this transformation
the ellipse with semi-axes $a_0$ and $b_0$ goes over to an ellipse
with semi-axes $a'_0 =a_0/\alpha $, $b'_0 =b_0/\beta $, while
any small line element, $dl$, directed at an angle $\varphi $
relative to the $x$ axis, is transformed into the element
   \begin{eqnarray} \nonumber
     dl'= dl \left({ \cos^2 \varphi \over \alpha^2} +
     { \sin^2 \varphi \over \beta^2} \right)^{1/2} \,,
   \end{eqnarray}
which is directed at an angle $\varphi'$ with
   \begin{eqnarray} \nonumber
     \tan \varphi'= {\alpha \over \beta} \tan \varphi \,.
   \end{eqnarray}
Since the current crossing the line element must be invariant, we
obtain the following transformation of the sheet current flowing
at an angle $\psi = \varphi \pm \pi/2$ relative to the $x$ axis:
   \begin{eqnarray} \nonumber
    J'( \psi') = J(\psi ) \beta \left[ 1 + \left({\alpha^2
    -\beta^2 \over \beta^2 } \right) \sin^2 \psi' \right]^{1/2} .
   \end{eqnarray}
 Thus, if the anisotropic critical current density in some elliptic
 sample with semi-axes $a_0'$, $b_0'$ can be approximated by the
 expression
   \begin{eqnarray}  %  32
    J'_c( \psi') = {\rm const}\cdot (1+\delta \sin^2 \psi')^{1/2} ,
   \end{eqnarray}
in which $\delta $ is some constant ($\delta > -1$), we
may reduce the critical state problem of this anisotropic
sample to the isotropic problem by performing the
transformation (31) from $x'$, $y'$ to $x$, $y$ with
$\beta =1$ and $\alpha = \sqrt{ 1 +\delta }$. After the
transformation we shall have an elliptic sample with semi-axes
$a_0= \sqrt{1 +\delta }\,a_0'$, $b_0 =b_0'$ and with isotropic
critical current $J_c=$ const,  and we can use the solution obtained
in Sec.\ II.

  Thus, our solution permits to analyze (at least qualitatively) the
features of the critical state of anisotropic superconducting films.
For example, for a circular disk with radius $R$ and with anisotropic
critical current obeying Eq.\ (32) with $\delta = 1$, one obtains
the transformed ellipse with semi-axes $a_0 = \sqrt2 R$,
$b_0 = R$ and the current distribution and stream lines depicted
in Figs.\ 4 - 7 for $e=0.7$, and the magnetic field depicted in
Figs.\ 9 - 10. These pictures then are transformed back to the
coordinates of the original circular disk by shrinking them
along $x$ by a factor of $0.7$.  Figure 12 shows another
example of a circular disk with larger anisotropy $\delta = 3$,
yielding $J_c(\pi/2) /J_c(0) = a_0/b_0 =  2$. The results
of Ref.\ \onlinecite{27} for square and rectangular plates with
anisotropic critical current are in agreement with this approach.
Our solution may also be used to include the in-plane anisotropy
of the critical current into theories \cite{14,28} which analyze
the critical state in flat superconductors taking into account
the current anisotropy (${\bf B}/B$ dependence) associated with
the curvature of the flux lines.

\section{Conclusion} % 5

   In this paper we derived an approximate analytic solution of
 the Bean critical state model for thin elliptic superconductors
 with semi-axes $a_0$ and $b_0=e a_0$  in a perpendicular
 field $H$. The accuracy of this solution is high
 and the known limiting cases of a circular disk and a long strip
 are recovered exactly. With increasing $H$ the penetrating flux
 front is an ellipse with axes $a$ and $b$ related by Eq.\ (13)
 such that at full penetration, when $b=0$, $a(b)$ stays finite,
 i.e., there is a section of length $2a(0)=2a_0 \cos(\pi e /2)$
 on the $x$ axis where regions with different orientation of the
 current meet, similar to the situation in rectangular
 platelets. \cite{9,11,27}  The current stream lines in the
 penetrated region deviate from
 ellipses and, in the fully penetrated state, they have a sharp
 bend on the $x$ axis in the interval $|x| \le a_0 - b_0^2/a_0$
 that is slightly wider than $2a(0)$.

    Our analytic solution shows that in non-circular finite
 superconductor films at any point not located near a symmetry axis
 or near the edge, when $H$ is increased the direction of the
 circulating sheet current changes until the flux front passes
 through this point, see Fig.\ 7. Therefore, in samples with
 thickness $d$ exceeding the London penetration depth $\lambda$,
 the orientation of the flux lines penetrating from the two
 flat surfaces changes also, since at the flat surfaces the
 flux lines are at a right angle to the sheet current.
 As a consequence, the flux-line arrangment rotates along $z$
 in the region where $B_z = 0$. Furthermore, it cannot be ruled
 out that also in the region with $B_z \ne 0$ the flux lines
 are twisted, i.e., their curvature is not planar.
 If this twist occurs, it means that the orientation of the
 current at fixed $x,y$ but different depth $z$ varies, and the
 integrated current density $J(x,y)$ is not necessarily equal
 to $j_c d$ or exactly constant even in the flux-penetrated region.
 The precise calculation of the flux and current distributions
 inside non-circular disks even of {\it small} thickness
 $d > \lambda$ then becomes an intricate 3D problem.
 The possibility of twisted flux lines in the penetrated region
 of non-circular thin disks should thus be considered in more detail
 in future work.

\acknowledgments
     We gratefully acknowledge stimulating discussions with
 John Gilchrist.

\appendix{ }
\section{Evaluation}

    The integrals (11) for $H/H_c$, (18) for $H_z(x,y)$, and (19)
 for $J_x$, $J_y$, and $g(x,y)$, like the elliptic integral of
 the second kind,
   \begin{eqnarray}  % A 1
    E(\varphi, k) = E(s, m) = s\int_0^1 \!\! \sqrt{1-ms^2t^2
    \over 1 -s^2t^2} \, dt
   \end{eqnarray}
 with $s = \sin\varphi$, $m=k^2$, may be evaluated with high precision
 by a substitution of variables. We first transform the integrals over
 $b'$ into integrals over a variable $t$ in the interval
 $0 \le t \le 1$, like Eq.\ (A1). Noting that these integrands may
 have poles $\propto 1/\sqrt{t}$ and $\propto 1/\sqrt{1-t}$ at the
 boundaries,  we use a substitution
   \begin{eqnarray}  % A 2
   I =\int_0^1 \!\! f(t)\, dt =\int_0^1 \!\! f[t(u)]\,t'(u)\,du
   \end{eqnarray}
 which has a weight function $t'(u) = dt/du$ that vanishes with a
 high power of $u$ and $1-u$ at $u=t=0$ and at $u=t=1$. A good such
 choice is
   \begin{eqnarray}  % A 3
    t  = g(u)  =35u^4 -84u^5 +70u^6 -20u^7\,, \nonumber \\
    t' = g'(u) =140 u^3 (1-u)^3 \,.
   \end{eqnarray}
 An even better choice is to iterate (A3) once, writing
   \begin{eqnarray}  % A 4
    t = g[g(u)] , ~~~ t' = g'(u)\, g'[g(u)] \,.
   \end{eqnarray}
 With (A4) one has $t \propto u^{16}$ and $t' \propto u^{15}$ near
 $t=0$, and $1-t \propto (1-u)^{16}$ and $t' \propto (1-u)^{15}$
 near $t=1$. The pole $f(t) \propto 1/\sqrt t$ in the original
 integrand is now removed in the substituted integrand
 $F(u) = f[t(u)]\, t'(u) \propto u^7$ and similarly near $t=1$.
 Since the new integrand vanishes rapidly at the boundaries of the
 integral, one may use an integration method with constant weights,
   \begin{eqnarray}  % A 5
    I = \int_0^1 \!\! F(u) \, du \approx {1\over N}\sum_{i=1}^N
    F\left( {i- 1/2 \over N} \right) \,.
   \end{eqnarray}
 This method achieves high precision $\approx 10^{-8}$ even with
 a small grid number $N= 20 \dots 30$.

   In Eqs.\ (11), (14), etc., $E(k)$ is given by
 the definition (A1) with $s=1$, and for $K(k)$ in Eq.\ (29)
 a similar definition applies. All these elliptic integrals are
 easily computed with high accuracy by this integration method.

 Before the integrals for $H_z(x,y)$ and $g(x,y)$ can be evaluated
 for a 2D grid of points $(x,y)$, one has to find the elliptic flux
 front that passed through the point $(x,y)$.
 This may be done by finding the zeros of the function
   \begin{eqnarray}  \nonumber % A 6
    p(b_{xy}) = {x^2\over a_{xy}^2} +{y^2\over b_{xy}^2} -1
   \end{eqnarray}
 with $a_{xy} = \cos[ b_0 \arccos (b_{xy}/b_0) ]$, cf.\ Eqs.\ (13)
 and (17) with length unit $a_0=1$. This is achieved by starting
 with $b_{xy} = b$ (the short semi-axis of the inner flux front)
 and iterating $ b_{xy} \leftarrow b_{xy} - 2\epsilon \, p(b_{xy})/
 [ p(b_{xy} +\epsilon) - p(b_{xy} -\epsilon) ]$ ($\epsilon \ll 1$)
 a few times.
 A similar procedure may be used to invert the relation $H(b)$,
 Eq.\ (11), to obtain $b(H)$. To obtain the correct $g(x,y)$
 also inside the flux front, and $H_z(x,y)$ also outside the
 superconductor, we chose $b_{xy} = b_0$ for all points outside
 the ellipse with semi-axes $a_0$, $b_0$, and $b_{xy} = b$ for all
 points inside the ellipse with semi-axes $a$, $b$ by putting
 $ b_{xy} \leftarrow {\rm max}[b, {\rm min} ( b_{xy}, b_0 )]$.

 %  \vspace{-0.5 cm}

  \vspace{\fill}
% \vspace{2.2 cm}
\epsfxsize= 1.00\hsize  \vskip 1.5\baselineskip
\centerline{ \epsffile{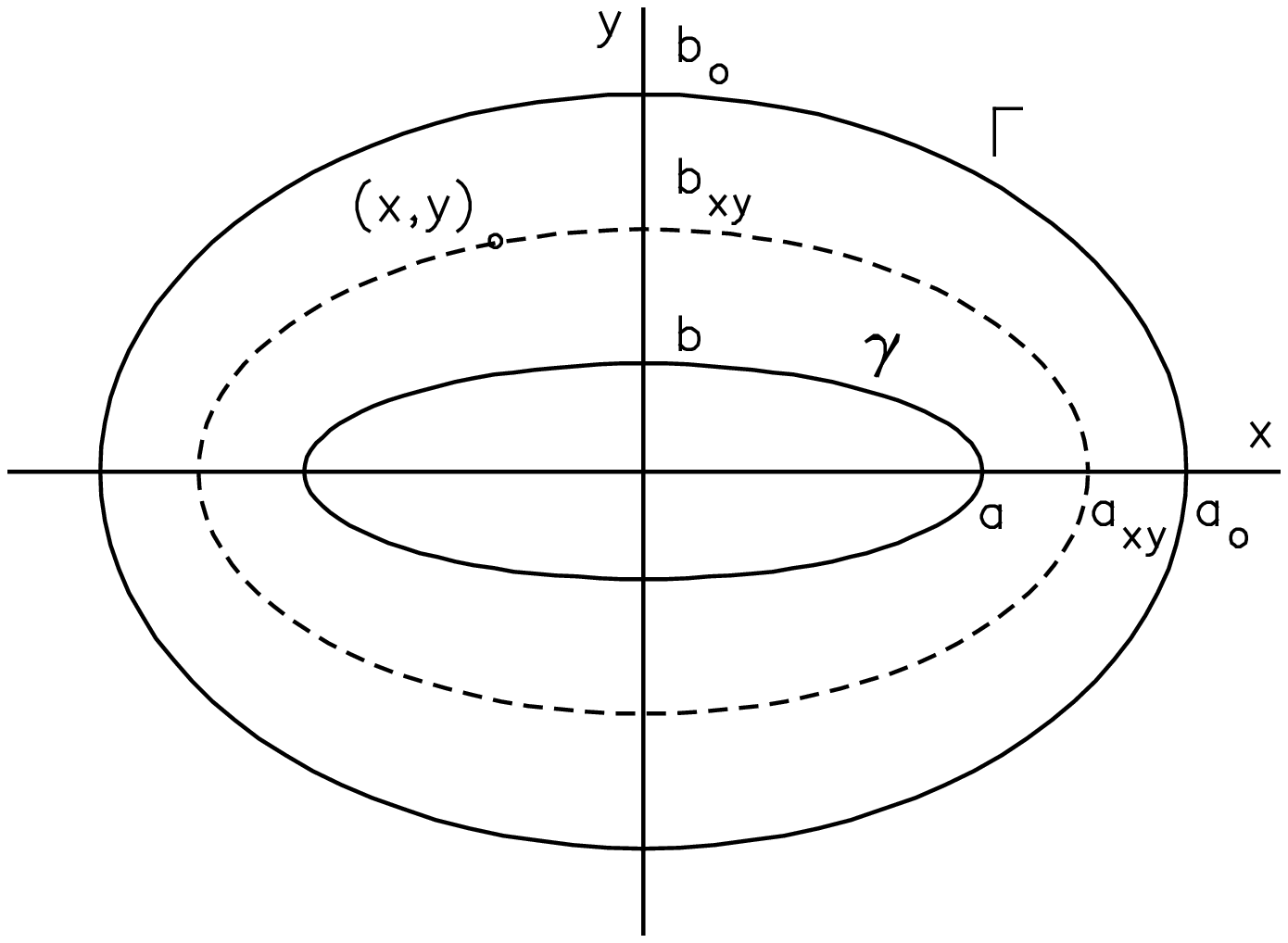}}
\begin{figure}[F1]
\caption{ The three ellipses defining the edge $\Gamma$ of the film 
 with semi-axes $a_0$, $b_0$; an earlier flux front passing through a
 given point $(x,y)$ (semi-axes $a_{xy}$, $b_{xy}$, dashed line);
 and the present flux front $\gamma$ with semi-axes $a$, $b$.
    }
\end{figure}

\epsfxsize= 1.00\hsize  \vskip 1.5\baselineskip
\centerline{ \epsffile{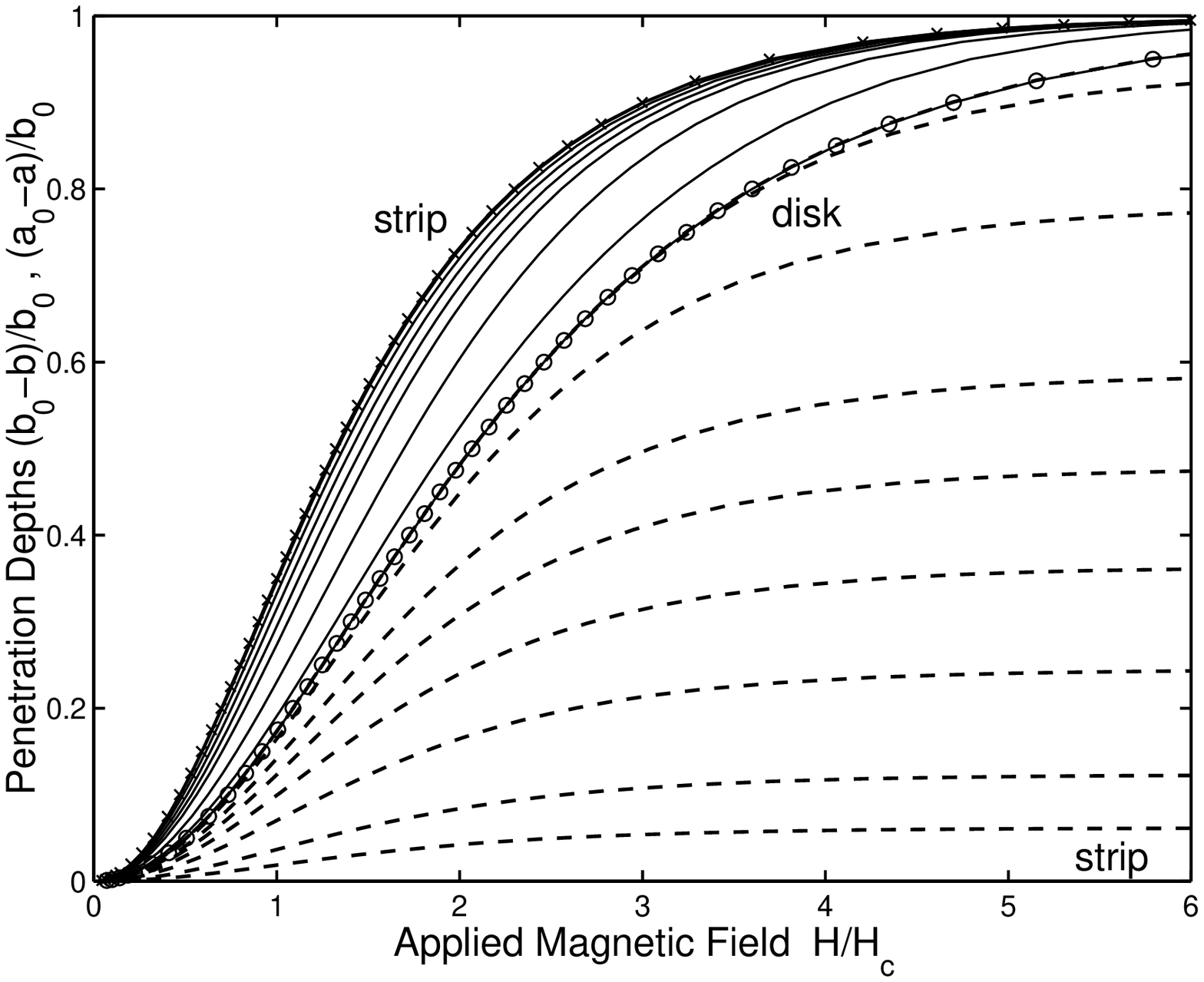}}
 \begin{figure}[F2]
\caption{ The semi-axes $a$ and $b$ of the elliptic flux fronts plotted
 as reduced penetration depths $(a_0 -a)/b_0$ (dashed lines) and
 $(b_0-b)/b_0$ (solid lines)  versus the applied field $H$
 (in units of $H_c = J_c/\pi$) for thin ellipses with excentricities
 $e=b_0/a_0 =$ 1, 0.9, 0.7, 0.5, 0.4, 0.3, 0.2, 0.1, 0.05, 0. The
 curve for the disk [$e=1$, $a=b$, Eq.\ (20) with $H_c$ replaced by
 $J_c/2$] is marked by circles. For strips, $b$ [Eq.\ (20)] is marked
 by crosses, and one has $a_0-a =0$.
    }
\end{figure}

\epsfxsize= 1.00\hsize  \vskip 1.5\baselineskip
\centerline{ \epsffile{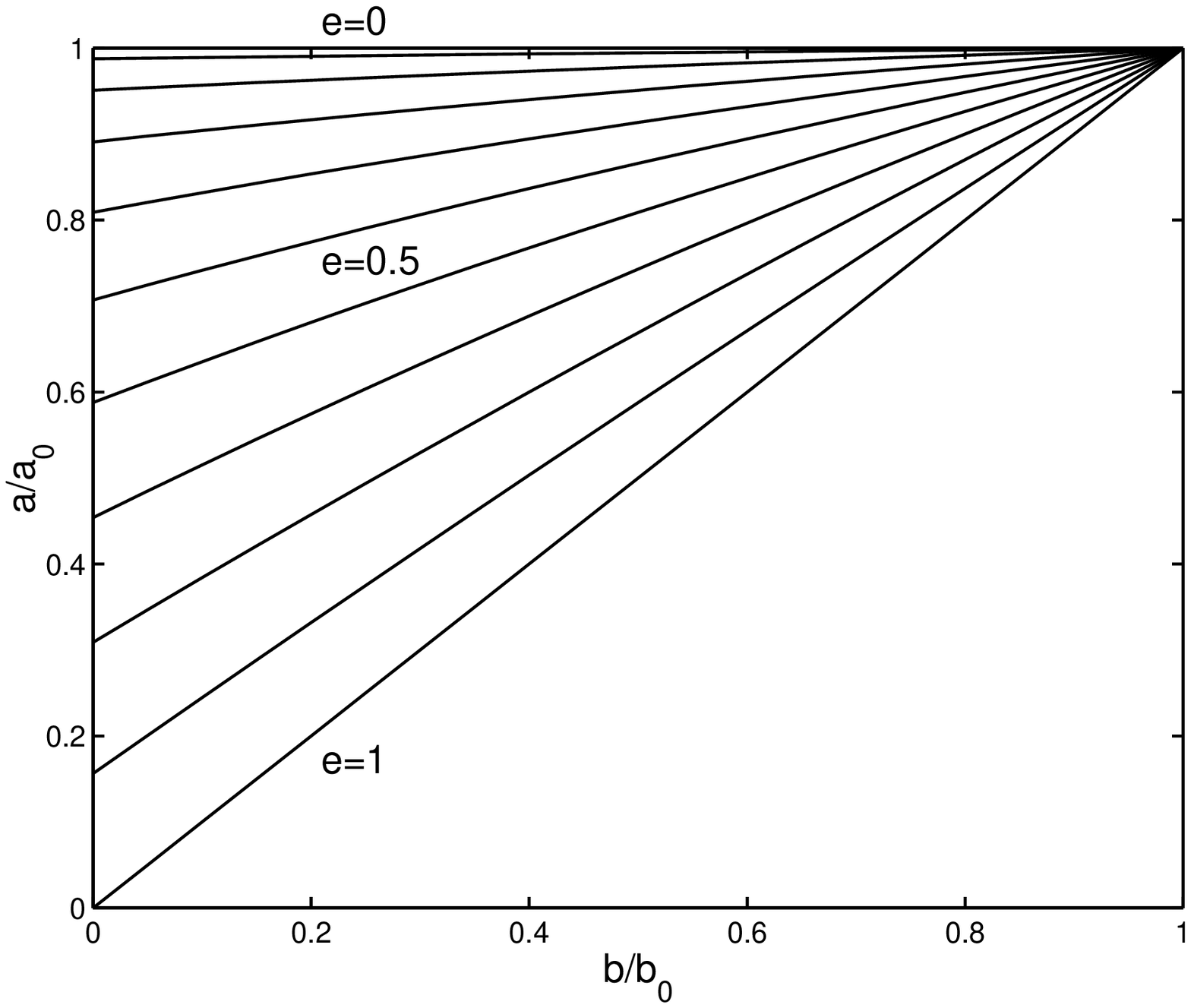}}
 \begin{figure}[F3]
\caption{ The long semi-axis $a$ as function of the short semi-axis
 $b$ of the penetrating elliptic flux front for various excentricities
 $e= b_0/a_0 =$ 1, 0.9, 0.8, \dots, 0.1, 0, Eq.\ (13).
    }
\end{figure}

\epsfxsize= 1.00\hsize  \vskip 1.5\baselineskip
\centerline{ \epsffile{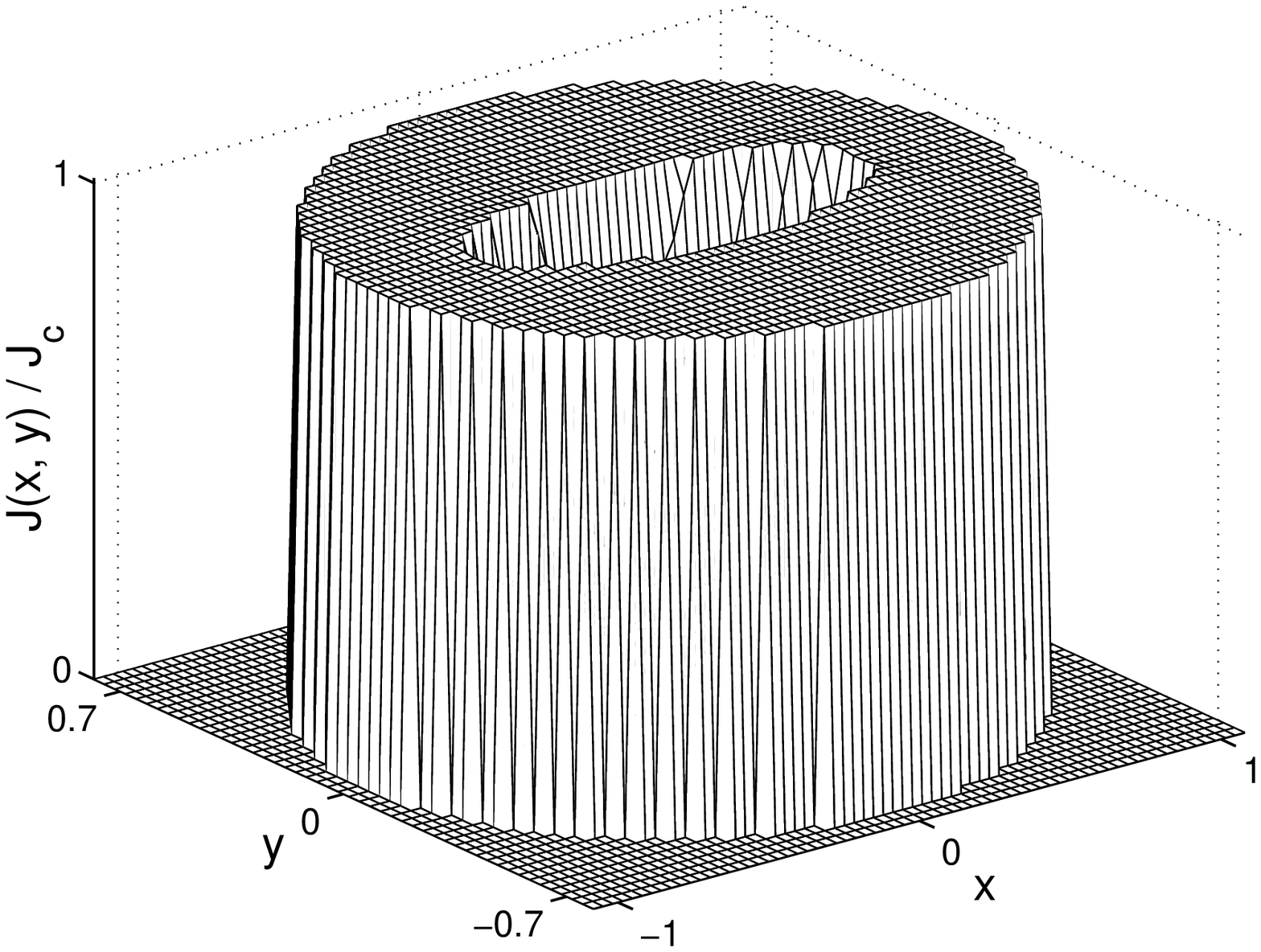}}
 \begin{figure}[F4]
\caption{ 3D plot of the magnitude of the sheet current
 $|{\bf J}(x,y)|$, Eq.\ (19), in a thin ellipse with $b_0/a_0 = 0.7$
 at $b/a_0=0.2$.
 In the flux-penetrated region the plotted $J/J_c$ deviates from
 unity by less than $10^{-3}$.
    }
\end{figure}

\epsfxsize= 1.00\hsize  \vskip 1.5\baselineskip
\centerline{ \epsffile{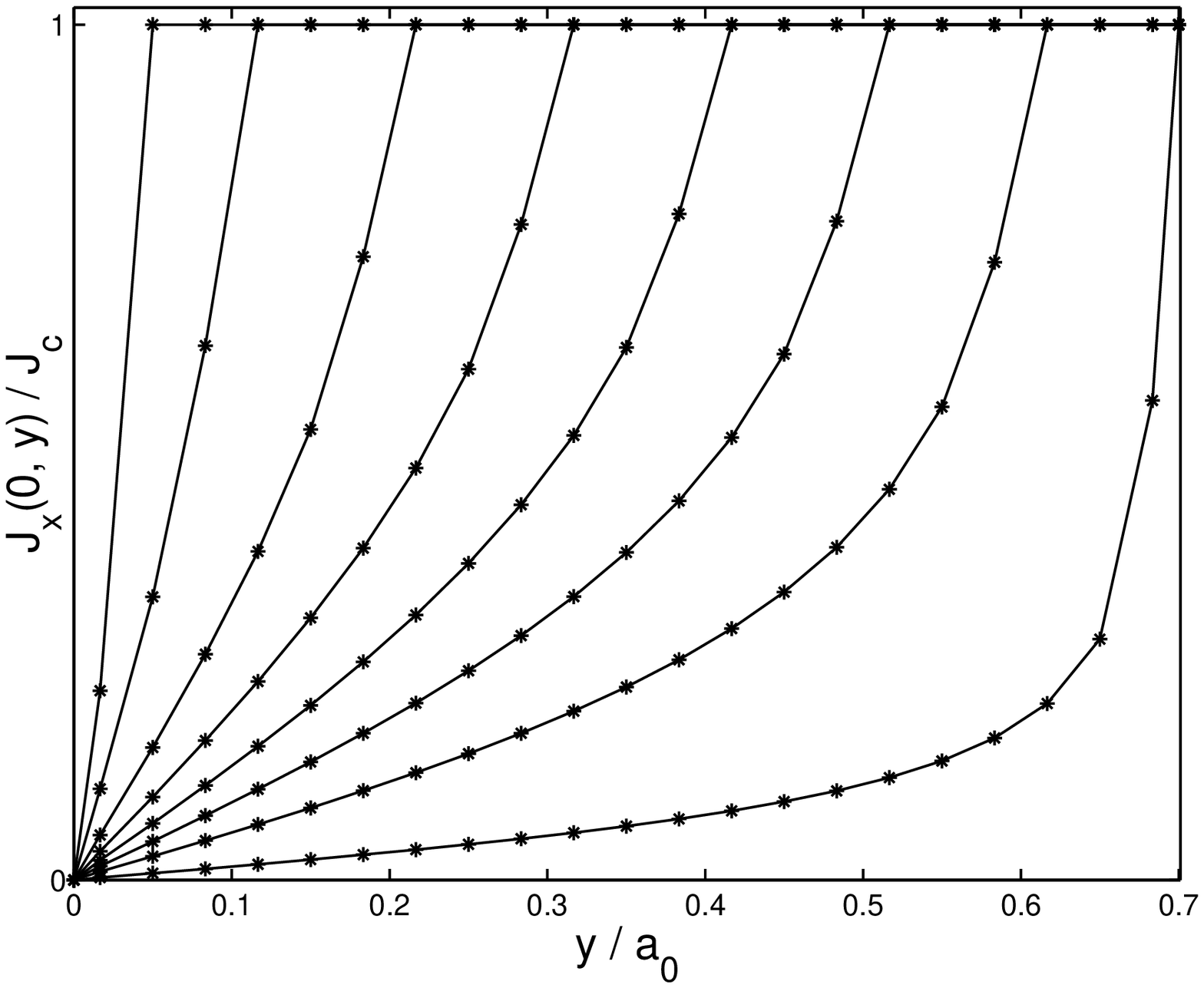}}
 \begin{figure}[F5]
\caption{ Profiles of the sheet current $J_x(0,y)$ in
 a thin ellipse with $b_0/a_0 =0.7$ for various degrees of flux
 penetration $b/a_0 =$ 0.69, 0.6, 0.5, \dots, 0.1, 0.05 (from right
 to left). Note the precise saturation to $J=J_c$, which demonstrates
 the accuracy of our approximation.
    }
\end{figure}

\epsfxsize= 1.00\hsize  \vskip 1.5\baselineskip
\centerline{ \epsffile{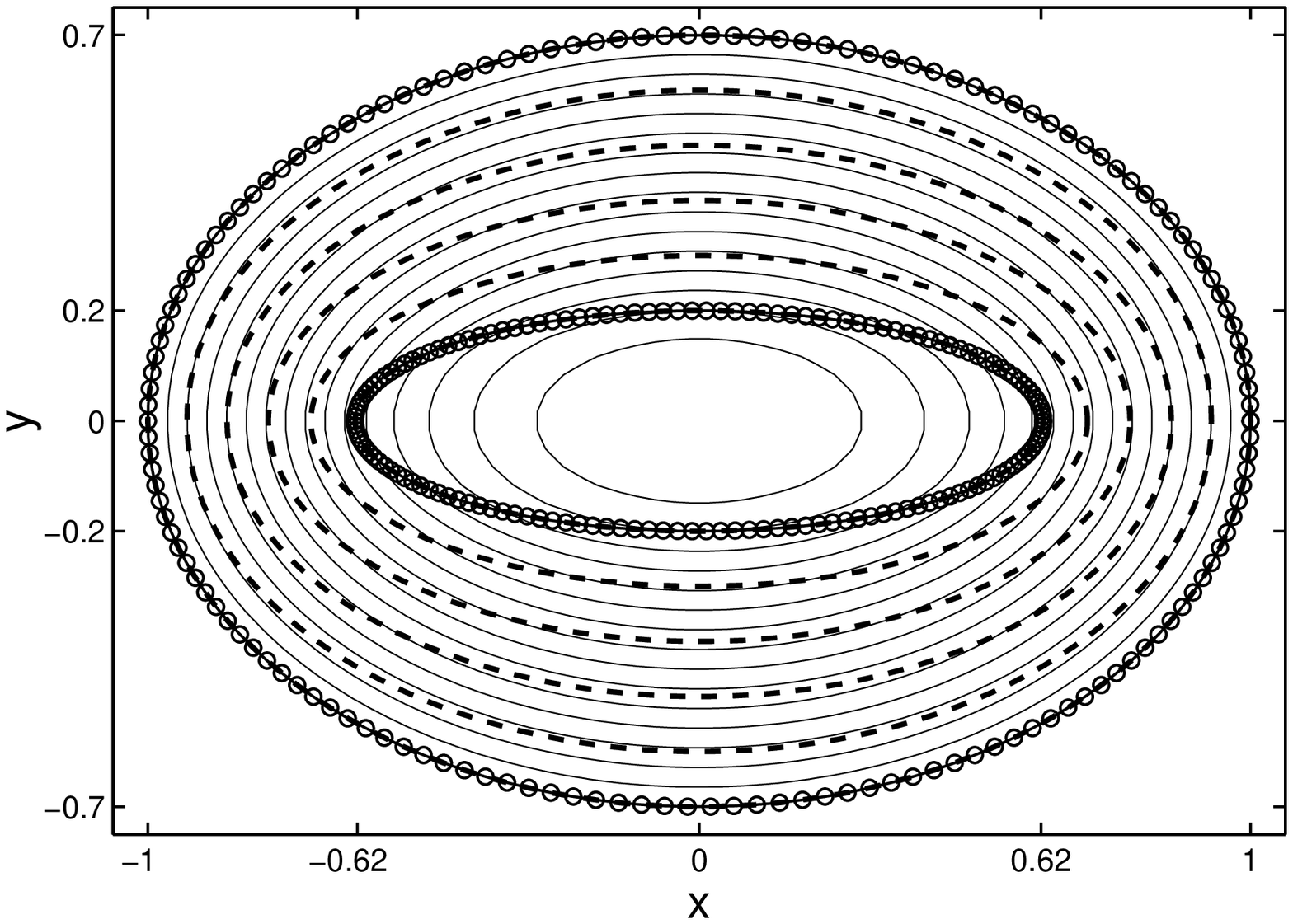}}
\epsfxsize= 1.00\hsize  \vskip 1.5\baselineskip
\centerline{ \epsffile{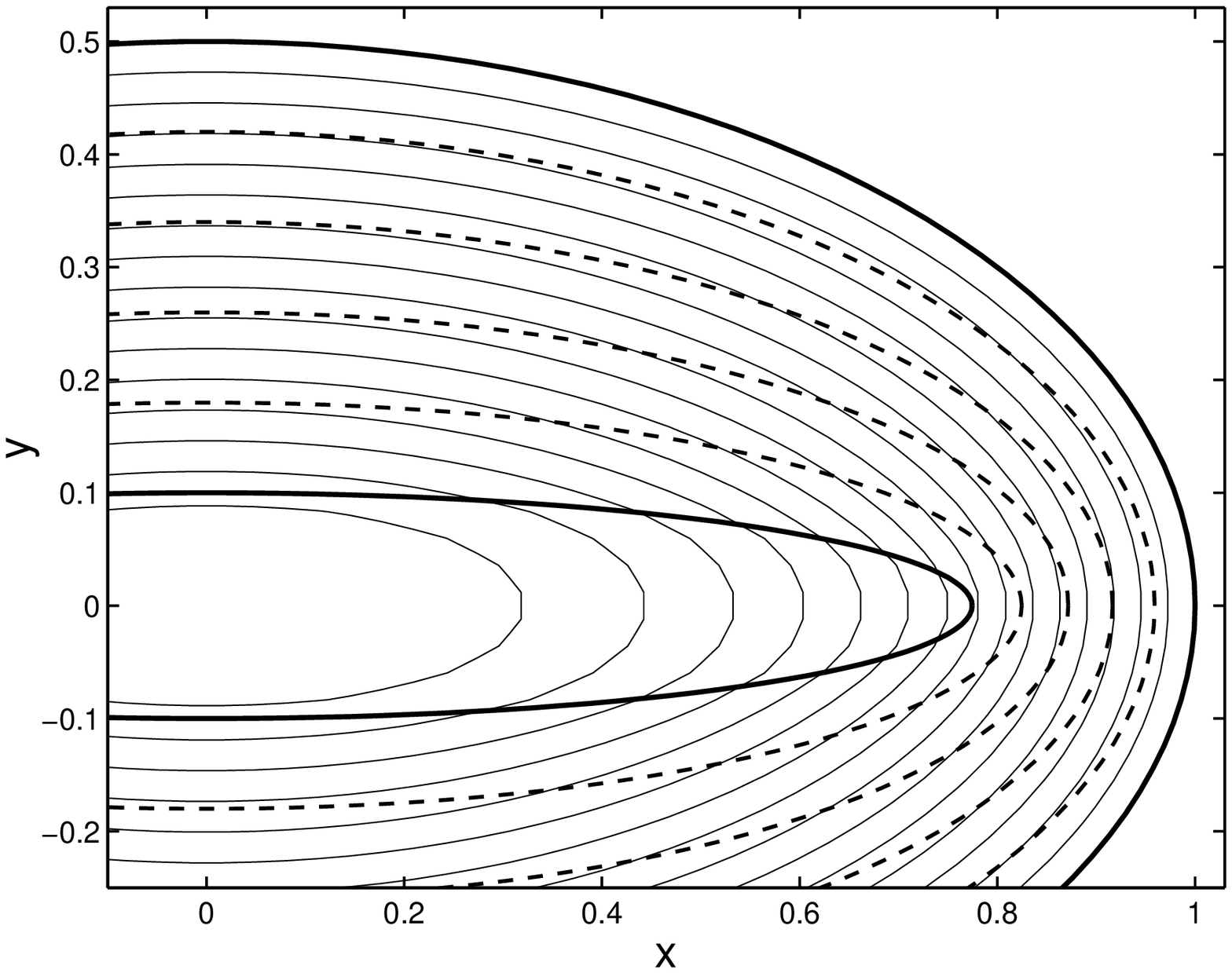}}
 \begin{figure}[F6]
\caption{ Stream lines (solid lines) of the sheet current
 ${\bf J}(x,y)$ in a thin  ellipse with (a) $b_0/a_0 = 0.7$,
 $b/a_0 =0.2$ (yielding $a= 0.624$) and (b) $b_0/a_0 = 0.5$,
 $b/a_0 = 0.1$ (yielding $a= 0.775$). Also shown are five previous
 flux fronts (dashed lines) at equidistant values of $b$.
 The edge of the film and the inner flux front are indicated by
 circles and a bold line. Note that in general the sheet current does
 not flow parallel to the flux fronts and that its magnitude is
 constant in the flux-penetrated region.
    }
\end{figure}

\epsfxsize= 1.00\hsize  \vskip 1.5\baselineskip
\centerline{ \epsffile{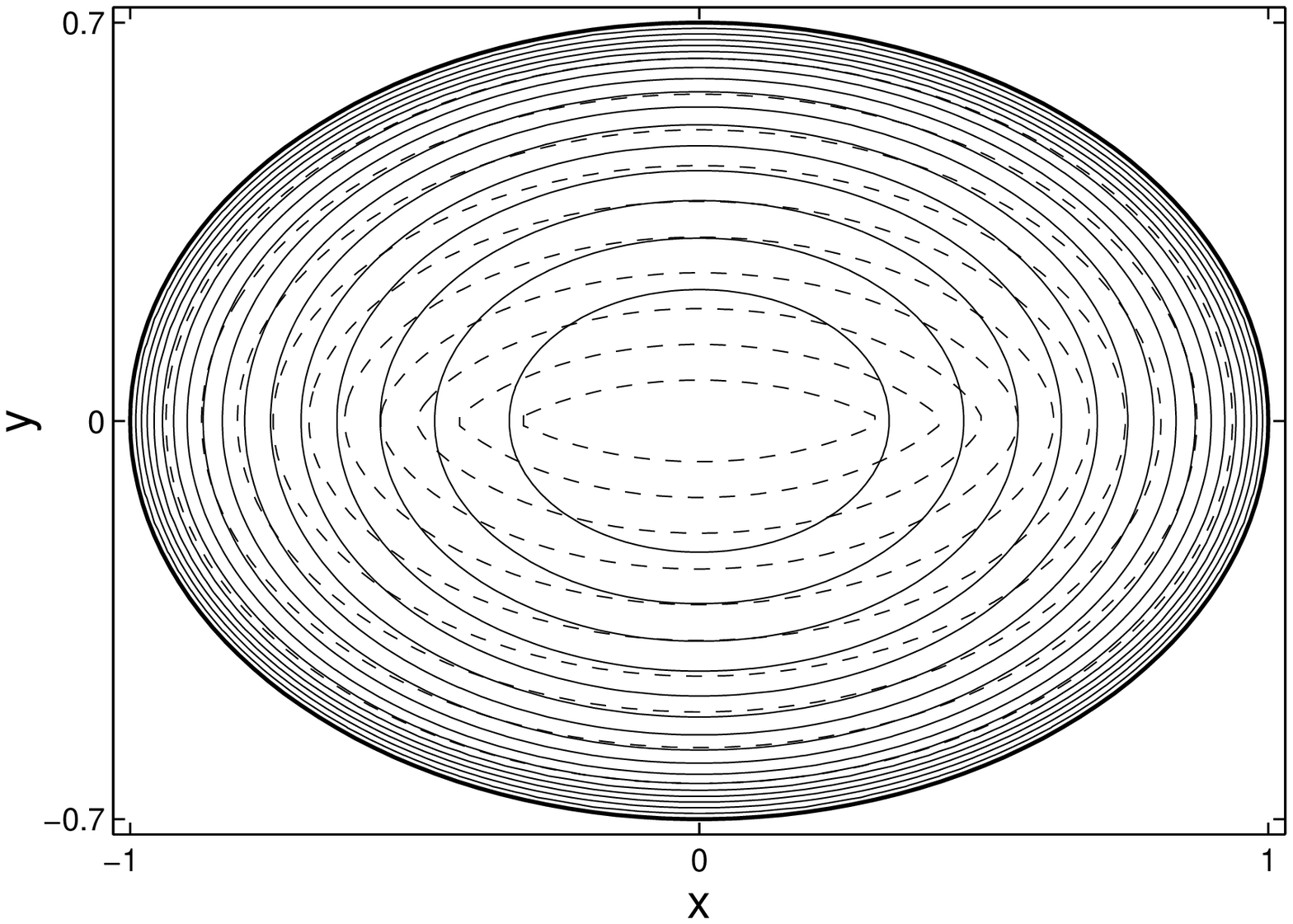}}
 \begin{figure}[F7]
\caption{ The stream lines of the sheet current in the same ellipse
 ($b_0/a_0=0.7$) at two different times during flux penetration: at
 $b/a_0=0.65$ (solid lines, almost ideal screening) and $b/a_0=0.01$
 (dashed lines, almost full penetration). Note that the orientation of
 the sheet current at some places changes strongly during the
 penetration, leading to a rotating vortex arrangment across
 the thickness of the film.
    }
\end{figure}

\epsfxsize= 1.00\hsize  \vskip 1.5\baselineskip
\centerline{ \epsffile{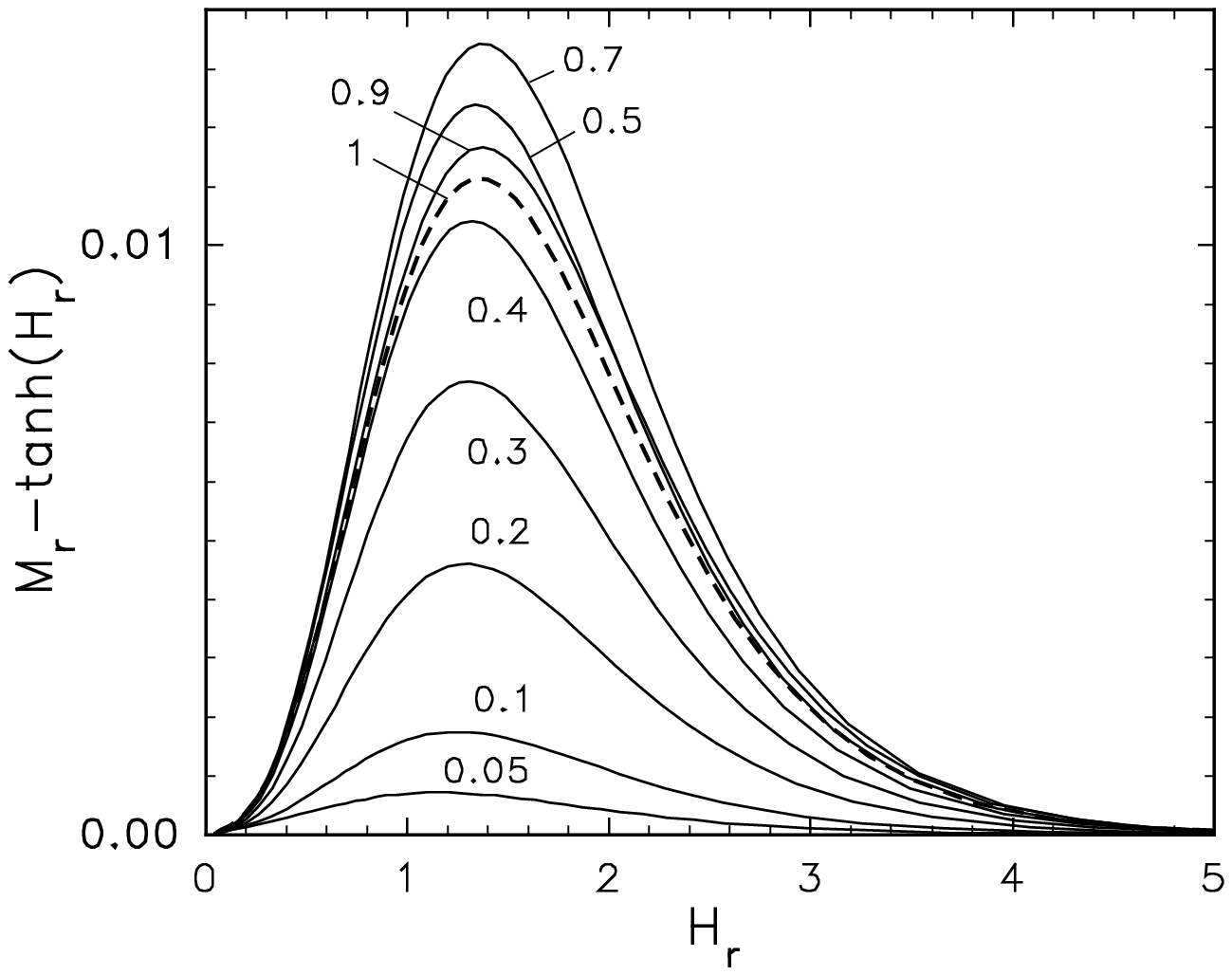}}
 \begin{figure}[F8]
\caption{ The magnetic moment (magnetization curve) of thin ellipses
 in a perpendicular magnetic field
 in the Bean model, plotted in normalized form as deviation from the
 strip result $M_r = \tanh(H_r)$ versus a reduced field $H_r=H/H_1$
 for excentricities $e=b_0/a_0 =$ 1, 0.9, 0.7, 0.5, 0.4, 0.3, 0.2,
 0.1, and 0.05, see Eqs.\ (11), (16), and (24) - (26).
    }
\end{figure}

\epsfxsize= 1.00\hsize  \vskip 1.5\baselineskip
\centerline{ \epsffile{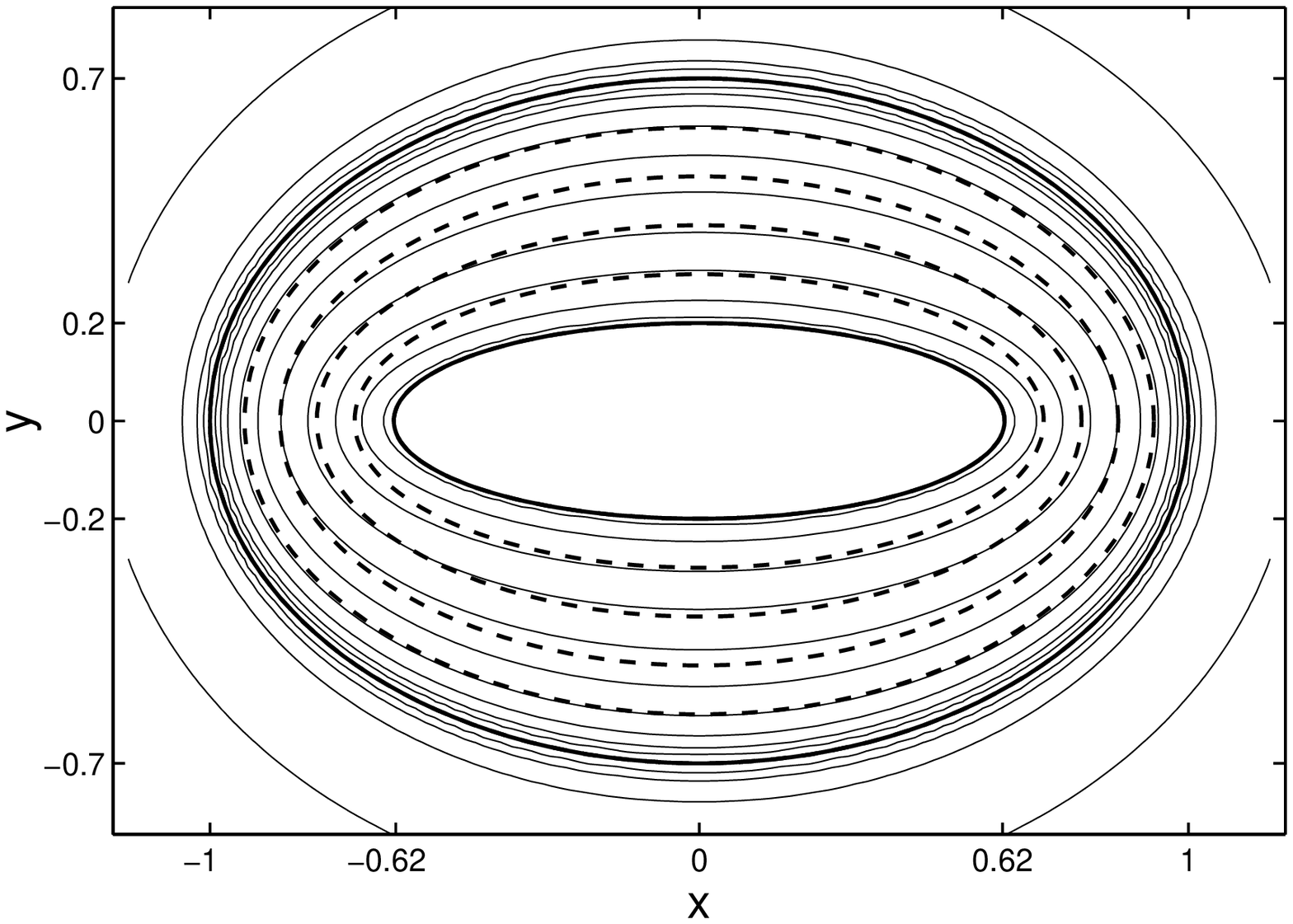}}
 \begin{figure}[F9]
\caption{ Contour lines (solid lines) of the perpendicular magnetic
 field $H_z(x,y)$ of a thin ellipse with $b_0/a_0 = 0.7$ in the plane
 $z=0$ for partial flux penetration with $b/a_0 = 0.2$, as in
 Fig.\ 6a. Note that these contours are nearly parallel to the
 elliptic flux fronts shown as dashed lines for $b/a_0 =$ 0.7, 0.6,
 0.5, 0.4, 0.3, and 0.2. The edge of the specimen and the inner
 flux front are marked by bold lines. Inside the inner flux front one
 has exactly $H_z(x,y) = 0$.
    }
\end{figure}

\epsfxsize= 1.00\hsize  \vskip 1.5\baselineskip
\centerline{ \epsffile{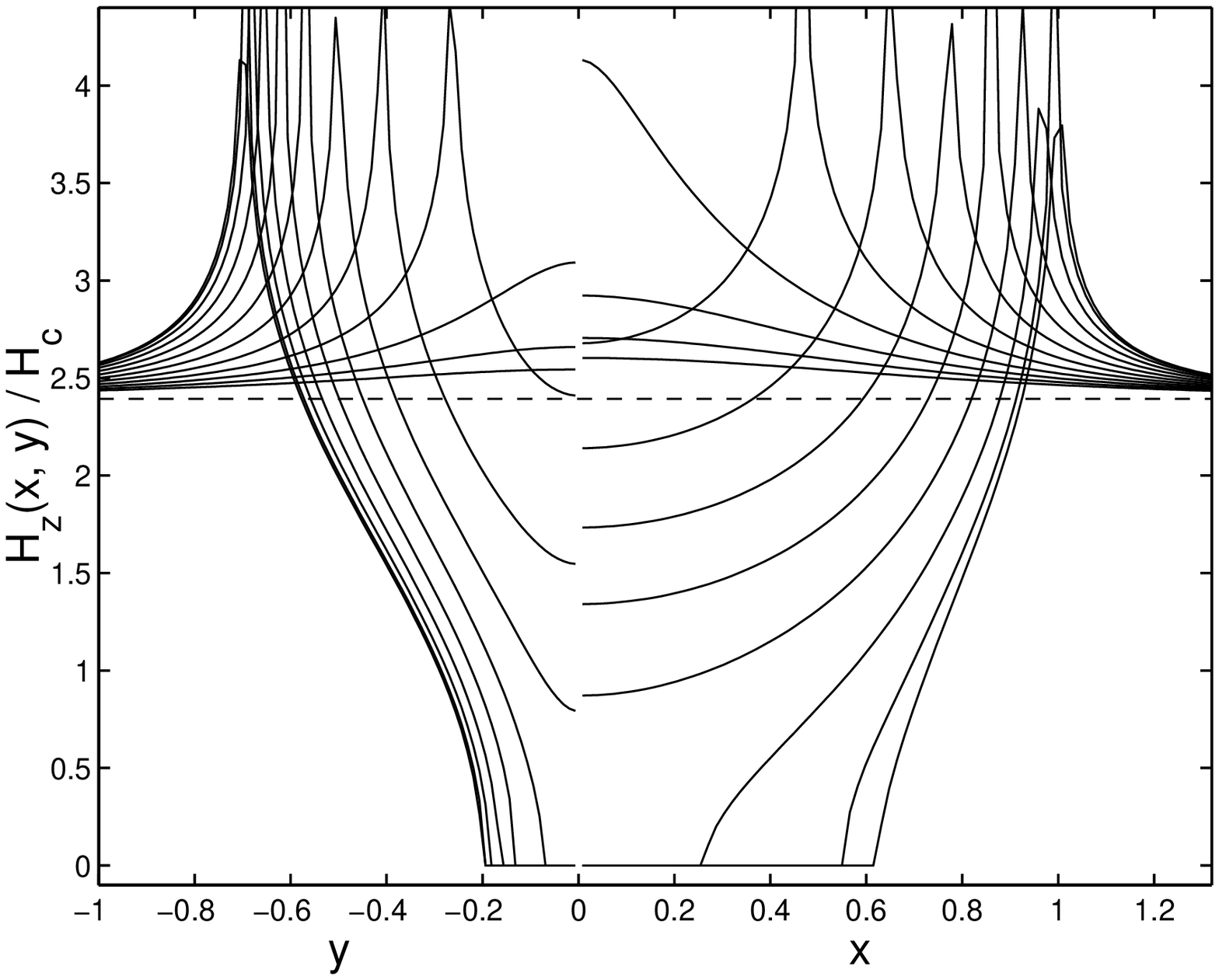}}
 \begin{figure}[F10]
\caption{ Profiles of the magnetic field $H_z(x,y)$ of Fig.\ 9
 plotted along $y$ (left) and along $x$ (right) for discrete
 equidistant values of $x$ or $y$, respectively. The dashed line
 marks the applied field $H/H_c=2.39$.
    }
\end{figure}

\epsfxsize= 1.00\hsize  \vskip 1.5\baselineskip
\centerline{ \epsffile{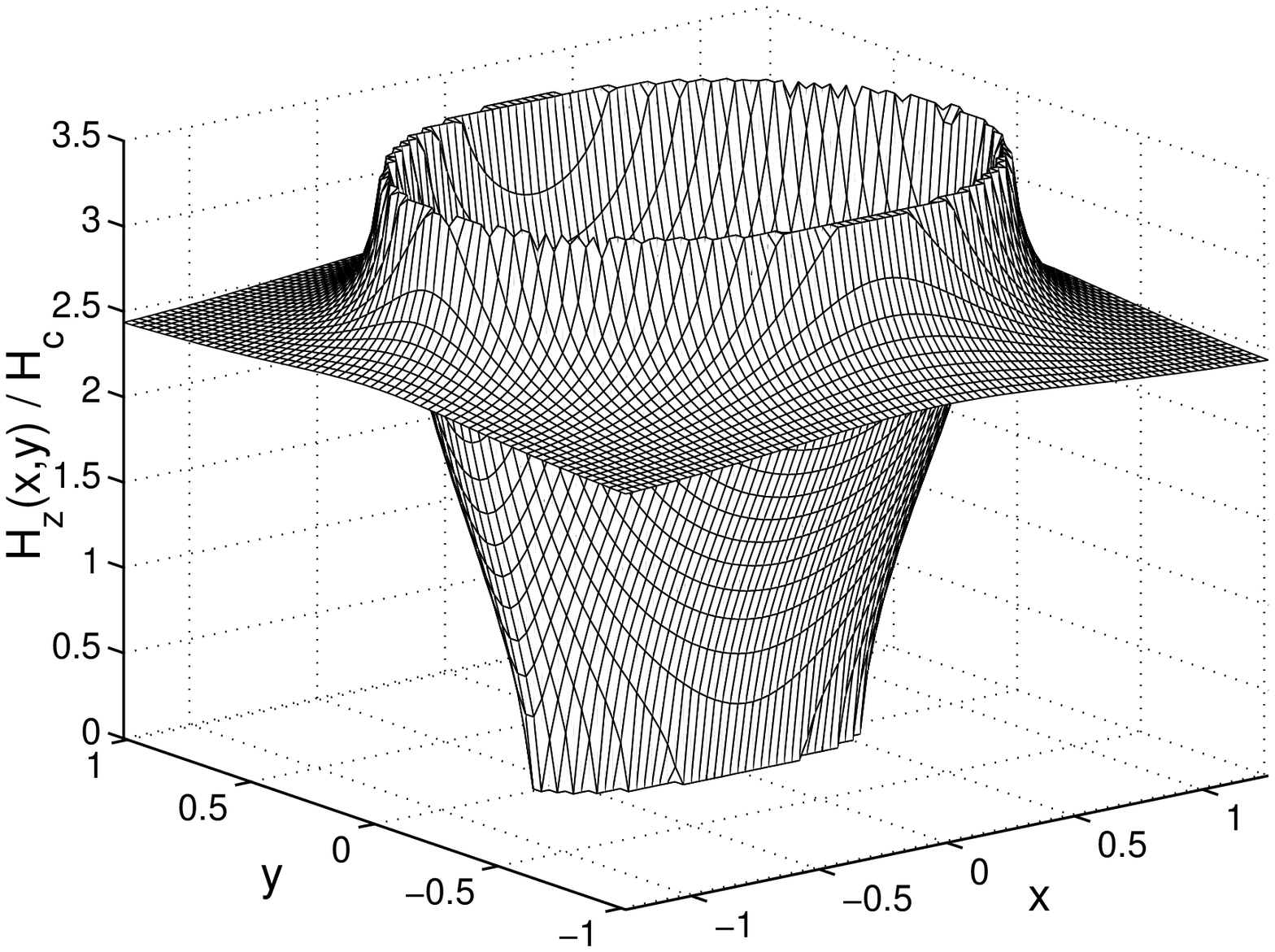}}
 \begin{figure}[F11]
\caption{ 3D plot of the magnetic field $H_z(x,y)$
 for $b_0/a_0=0.7$ and $b/a_0=0.2$ as in Figs.\ 9 and 10. The
 logarithmic infinity at the edge of the ellipse is cut off at
 $H_z/H_c = 3.5$ to limit the irregular peaks caused by the
 equidistant grid used for this plot.
     }
\end{figure}

\epsfxsize= 1.00\hsize  \vskip 1.5\baselineskip
\centerline{ \epsffile{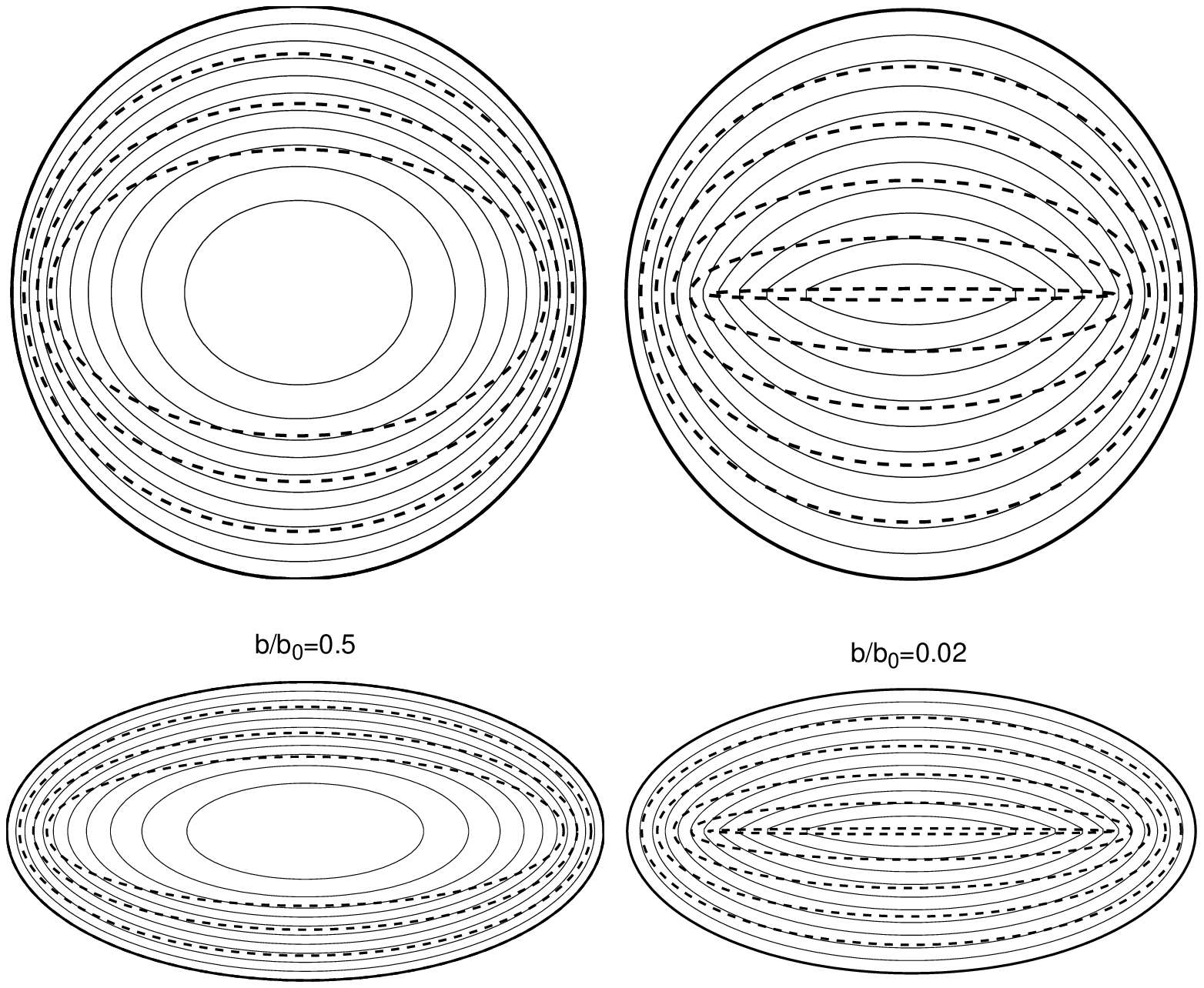}}
 \begin{figure}[F12]
\caption{ Current stream lines (solid curves) and flux fronts
 (dashed curves) in a circular disk with anisotropic critical
 current density,  $J_{cy}/J_{cx} = J(\pi/2)/J_c(0) = 2$
 [$\delta = 3$ in Eq.\ (32)] for half ($b/b_0=0.5$, left) and
 full ($b/b_0 = 0.02$, right)  penetration of flux.
 Shown at the bottom is the isotropic ellipse with axis ratio
 $b_0/a_0 =0.5$ to which this problem transforms.
 This ellipse exhibits equidistant current stream lines outside
 the inner flux front.
    }
\end{figure}

 \end{multicols}
 \end{document}